\documentclass[conference]{IEEEtran}
\IEEEoverridecommandlockouts
\usepackage{hyperref}
\hypersetup{colorlinks=true, citecolor=blue, linkcolor=blue, urlcolor=blue}
\usepackage{cite}
\usepackage{amsmath,amssymb,amsfonts}
\usepackage{graphicx}
\usepackage{tabularx}
\usepackage{textcomp}
\usepackage{xcolor}
\usepackage{xspace}
\usepackage{algorithm}
\usepackage{algpseudocode}
\usepackage[a4paper, total={184mm,239mm}]{geometry}
\def\BibTeX{{\rm B\kern-.05em{\sc i\kern-.025em b}\kern-.08em
    T\kern-.1667em\lower.7ex\hbox{E}\kern-.125emX}}
\usepackage{comment}

\newcommand{\Menage}{\textsc{Menage}\xspace}
\newcommand{\PE}{\textsc{MX-NeuraCore}\xspace}
\newcommand{\Syn}{\textsc{A-Syn}\xspace}
\newcommand{\NE}{\textsc{A-Neuron}\xspace}
\newcommand{\memone}{\textsc{MEM\textsubscript{E}}\xspace}
\newcommand{\memtwo}{\textsc{MEM\textsubscript{E2A}}\xspace}
\newcommand{\memthree}{\textsc{MEM\textsubscript{S\&N}}\xspace}

\newcommand{\myfigref}[1]{\hyperref[#1]{Figure \ref*{#1}}}
\newcommand{\mytabref}[1]{\hyperref[#1]{Table \ref*{#1}}}
\newcommand{\myalgref}[1]{\hyperref[#1]{Algorithm \ref*{#1}}}
\newcommand{\myeqref}[1]{\hyperref[#1]{(\ref*{#1})}}

\makeatletter
\renewcommand{\@cite}[2]{\textcolor{blue}{[#1\if@tempswa , #2\fi]}}
\makeatother

\begin{document}


\title{\Menage: Mixed-Signal Event-Driven Neuromorphic Accelerator for Edge Applications\\
}

\author{
    Armin Abdollahi, Mehdi Kamal, and Massoud Pedram\\
    Department of Electrical and Computer Engineering, University of Southern California, Los Angeles, CA, USA\\
    \{arminabd, mehdi.kamal, pedram\}@usc.edu
}

\maketitle

\begin{abstract}
This paper presents a mixed-signal neuromorphic accelerator architecture designed for accelerating inference with event-based neural network models. This fully CMOS-compatible accelerator utilizes analog computing to emulate synapse and neuron operations. A C2C ladder structure implements synapses, while operational amplifiers (op-amps) are used to realize neuron functions.
To enhance hardware resource utilization and power efficiency, we introduce the concept of a virtual neuron, where a single neuron engine emulates a set of model neurons, leveraging the sparsity inherent in event-based neuromorphic systems. 
Additionally, we propose a memory-based control technique to manage events in each layer, which improves performance while maintaining the flexibility to support various layer types. We also introduce an integer linear programming (ILP)-based mapping approach for efficiently allocating the model onto the proposed accelerator.
The accelerator is a general-purpose neuromorphic platform capable of executing linear and convolutional neural models. The effectiveness of the proposed architecture is evaluated using two specially designed neuromorphic accelerators and two event-based datasets. The results show that the proposed architecture achieves 12.1 TOPS/W energy efficiency when accelerating a model trained on CIFAR10-DVS.
\end{abstract}


\section{Introduction}

Neuromorphic computing, inspired by the architecture and functionality of the human brain, offers a highly efficient alternative to traditional computing paradigms \cite{ref1}.  In the brain, neurons communicate through electrical impulses or spikes, which allow for parallel processing and event-driven computation \cite{ref3}. Spiking Neural Networks (SNNs) are a key component of neuromorphic computing, utilizing discrete spikes for communication between artificial neurons \cite{ref5}. This spike-based approach contrasts with traditional artificial neural networks (ANNs) that rely on continuous values, leading to significant reductions in power consumption and enabling the natural processing of temporal information \cite{ref6}. Moreover, SNNs have shown potential in various applications, including sensory processing, autonomous systems, and robotics, where low latency and real-time responses are critical. SNN also could be used on medical diagnostics \cite{ref44}, and hardware optimization \cite{ref43, ref45} to enhance the efficiency. Additionally, The use of spiking neural networks (SNNs) could complement traditional deep learning models in mortality prediction \cite{ref41, ref42} by leveraging the event-driven processing of temporal health data to improve efficiency. By leveraging the sparse nature of spike-based communication, SNNs excel in environments where data is dynamic and computational efficiency is essential, making them ideal for edge computing and other low-power scenarios \cite{ref40}.

The Leaky Integrate-and-Fire (LIF) neuron model, commonly used in SNNs, emulates the behavior of biological neurons by accumulating input signals and firing spikes when a threshold is exceeded, making SNNs suitable for tasks involving time-dependent data \cite{ref7}. This model not only provides a biologically plausible representation of neural activity but also introduces inherent noise robustness and adaptability to varying input patterns, enhancing the system's ability to handle complex, time-varying stimuli. The LIF model’s reliance on simple arithmetic operations further contributes to the computational efficiency of SNNs, making them a promising alternative for neuromorphic processors.

Despite these advantages, scaling neuromorphic systems remains a significant challenge due to hardware constraints and energy inefficiencies. One critical limitation of existing neuromorphic chips is their high energy consumption and hardware complexity when scaling to large datasets and networks \cite{ref9}. Traditional SNN implementations are often computationally expensive due to dense synaptic connections and the need for high precision in neural computations.
One promising solution for improving energy efficiency is analog and mixed-signal computing. They can significantly enhance energy efficiency by directly processing in analog form, which reduces the need for power-intensive digital conversions, making them particularly suitable for handling the sparse, event-driven nature of neural models in neuromorphic applications.

Memristor-based neuromorphic systems, which have gained attention for their potential to enable efficient in-memory computing, also face significant challenges such as device variability, limited endurance, and resistance drift, which can severely impact the stability and accuracy of neural computations \cite{ref31, ref32}. These issues make it difficult to maintain consistent performance, particularly in large-scale deployments where variability can lead to unpredictable behavior. As neuromorphic systems become increasingly deployed in real-time and resource-constrained environments, reducing power consumption while maintaining accuracy is paramount.

To address the aforesaid challenges, this paper describes a fully CMOS-compatible mixed-signal neuromorphic accelerator architecture, which leverages analog computing to emulate the synaptic connections and neuron behavior. 
The C2C ladder structure, which can act as a high-precision analog multiplier, is used to scale the spikes based on the model parameters, and op-amp-based neurons have been used to emulate the LIF behavior.
To improve the energy efficiency and because of the sparsity in the events, we suggest modeling more than one neuron in each physically designed neuron engine. 
Moreover, to fully utilize the accelerator components, we suggest an integer linear programming (ILP) formulation to mathematically formulate the mapping problem and solve it efficiently. 
The effectiveness of the designed accelerator is assessed on two well-known neuromorphic datasets.



The remainder of this paper is organized as follows. Section II reviews the prior work. The details of the proposed accelerator architecture and the ILP formulation for the mapping are discussed in Section III. Section IV presents the evaluation results, and finally, the paper is concluded in Section V.

\section{Related Work}

Neuromorphic hardware, such as IBM's TrueNorth and Intel's Loihi, is designed to support the unique requirements of SNNs, with architectures optimized for sparse, event-driven computation \cite{ref8, ref9}. These chips integrate multiple cores to simulate vast networks of neurons and synapses efficiently, enabling real-time processing while maintaining low power consumption. \cite{ref12} demonstrates the use of memristor-based neural networks for efficient in-situ learning with adaptive capabilities to hardware imperfections, achieving competitive classification accuracy on standard machine learning datasets. In \cite{ref13}, fully memristive neural networks are implemented using memristors for both neurons and synapses, facilitating pattern classification via unsupervised synaptic weight updates. \cite{ref14} explores the cooperative development of memristor-based spiking neural networks (SNNs), emphasizing the integration of neural network architectures and memristor technology for efficient energy-saving systems. \cite{ref15} discusses the BrainScaleS-2 accelerated analog neuromorphic computing architecture, which integrates both analog and digital components for efficient spike-based processing. In \cite{ref16}, the authors present advancements in CMOS-integrated memristive arrays, emphasizing their use in high-performance computing systems based on brain-inspired architectures. The paper discusses how memristors, combined with crossbar architectures, enable efficient in-memory computing for vector-matrix multiplication. 
In \cite{ref17}, large-scale memristor crossbar arrays are implemented for neuromorphic computing, enabling efficient parallel in-memory processing for neural networks. \cite{ref18} introduces a sparsity-aware neuromorphic computing unit that combines spiking and artificial neural networks using leaky integrated neurons. \cite{ref20} proposes a stochastic-bits enabled binary spiking neural network (sBSNN) that uses probabilistic computing for energy-efficient neuromorphic systems. The network leverages binary synapses and stochastic neuron. In \cite{ref21}, the authors present NeuRRAM, a compute-in-memory chip utilizing resistive memory for energy-efficient edge AI applications.
\cite{ref22} presents a 4-Mbyte compute-in-memory (CIM) macro based on resistive random-access memory (ReRAM) for AI edge devices. \cite{ref23} present a 4M-synapse integrated neural network processor based on analog ReRAM, with cell current-controlled writing to achieve high power efficiency. \cite{ref24} introduces a fully integrated analog ReRAM-based compute-in-memory chip for efficient neural network processing with parallel MAC operations. \cite{ref25} presents a mixed-signal spiking neural network processor with 8-bit synaptic weights and on-chip learning that operates in the continuous-time domain. \cite{ref26} propose a clock-free spiking neural network for AIoT applications, using a multilevel event-driven architecture to reduce power consumption and inference latency. \cite{ref27} shows a fully integrated spiking neural network combining analog neurons and resistive RAM (RRAM) synapses, tailored for the N-MNIST classification task. 

\section{\Menage Architecture}
\begin{figure}[ht]
    \centering
    \includegraphics[width=1\linewidth]{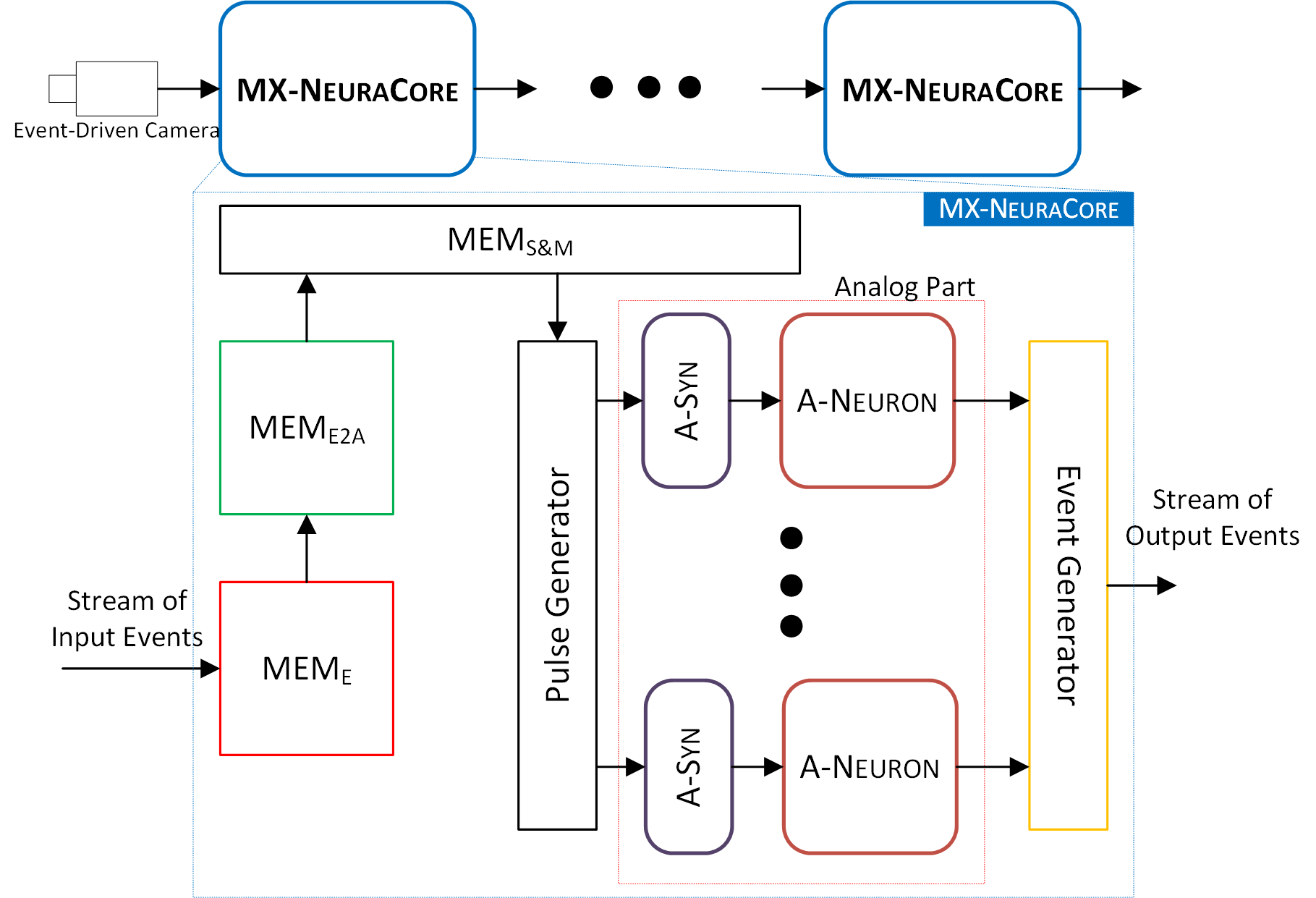} 
    \caption{The proposed \Menage architecture}
    \label{fig:whole} 
\end{figure}

The general structure of the proposed neuromorphic accelerator architecture (called \Menage) is depicted in \myfigref{fig:whole}. 
The architecture contains a chain of mixed-signal Neuromorphic (\PE) engines, each used for executing one layer of the given neural model. Each \PE consists of a memory-based controller to manage the received events to the layer as well as scheduling and sending them to the analog synapse (\Syn) engines. The output of \Syn is connected to the Analog Neuron (\NE) engine. The generated pulse by an \NE of a \PE is passed to the next \PE. 
In the proposed architecture, the weights are mapped to the SRAM memories of {\Syn}s, while control signals generated by a distiller are stored on the memories of the proposed memory-based controller. The control signals are generated based on the proposed ILP-based mapping approach, which aims to improve the utilization of the \Syn and \NE engines inside each \PE.
The following subsections will discuss the details of the engines and units of the \PE. 

The proposed accelerator supports rate-based spike encoding where spikes are pulses passed between the {\PE}s. The weights are in the 8-bit digital format. Hence, the parameters of the given model should be quantized before mapping them to the accelerator's memory. In addition, since our proposed accelerator supports pruned neural models, we suggest pruning the network before mapping the model onto the accelerator. The general flow for mapping the given model on the proposed accelerator is illustrated in \myalgref{alg:network}.

\begin{algorithm}[htbp]
\caption{Training, Pruning, Quantization, and Mapping Process}
\label{alg:network}
\begin{algorithmic}[1]
\State \textbf{Input:} Dataset, Neural model, and Accelerator specifications 

\State \textbf{Step 1: Train Network}
\State Train the network on a dataset.
  
\State \textbf{Step 2: Pruning and Quantization}
\State Apply pruning to reduce the number of synaptic connections in the network.
\State Apply quantization to reduce the precision of synaptic weights.
\State Ensure the network fits within the constraints of the hardware architecture.

\State \textbf{Step 3: Extract Weights and Spikes}
\State Extract the pruned and quantized weights from the trained network.

\State \textbf{Step 4: Generate the ILP Formulation}
\State Formulate the Integer Linear Programming (ILP) problem based on the extracted weights for each layer in different time steps.

\State Define the objective function and constraints to mimic the Python-level spiking neural network behavior.

\State \textbf{Step 5: Mapping and Executing on the Hardware}
\State Store the ILP solutions as the config bits on the internal memories.
\State Store the quantized parameters on the weight memory of  {\Syn}s.
\State Using LIF Neurons to generate the output spikes for different layers.
\State Determining the output class based on the output spikes.

\State \textbf{End}
\end{algorithmic}
\end{algorithm}

A system clock is used to synchronize the digital components. When the clock's rising edge occurs, any new event received by a \PE is stored in the Event Memory (\memone). Each received event contains the index of the source neuron. The \PE includes a controller, which utilizes a polling-based approach to check the \memone in each clock cycle. In the case of an event, the controller extracts the initial address and the number of the rows inside the Synapse and Neuron Assignment memory (\memthree), which contains the synapse connections. This information is obtained via an Event-to-Address memory (\memtwo). It may take more than one clock cycle to dispatch the received event from a neuron to the destination neurons. In this case, the controller does not fetch any new event from the \memone.


In \memtwo, the initial bits specify the number of connections or rows that need to be read from \memthree for a specific element in \memtwo. In contrast, the subsequent bits provide the starting address of these connections in \memthree. This sequential storage scheme facilitates efficient access and processing of synaptic connections. The \memthree utilizes an assignment derived from the Integer Linear Programming (ILP) solution, which allocates each connection to the appropriate \PE hardware component, optimizing resource utilization and ensuring adherence to hardware constraints.


\subsection{\NE Structure}

\begin{figure}[ht]
    \centering
    \includegraphics[width=0.7\linewidth]{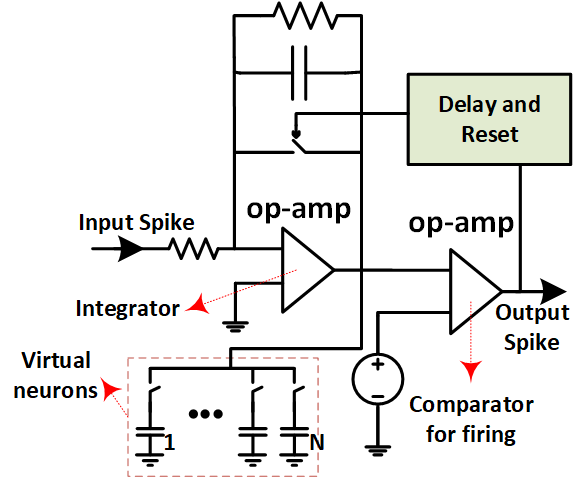} 
    \caption{\NE architecture.}
    \label{fig:LIF} 
    \vspace{-1.5 em}
\end{figure}

The \NE emulates the leaky integrate-and-fire (LIF) neuron, whose structure is illustrated in \myfigref{fig:LIF}. In the LIF neuron, the membrane potential \(V(t)\) of the neuron evolves according to the differential equation as shown in \myeqref{eq:leaky}.

\begin{equation}
\tau_m \frac{dV(t)}{dt} = -V(t) + R_m I(t)
\label{eq:leaky}
\end{equation}

In the above equation,
\( \tau_m \) denotes the membrane time constant,
\( V(t) \) is the membrane potential at time \( t \),
\( R_m \) represents the membrane resistance,
\( I(t) \) is the input current to the neuron. A neuron fires a spike when \( V(t) \) exceeds a threshold \(V_{th}\), and the membrane potential is reset to \( V_{reset} \). The implemented neuron emulates the LIF neuron model using discrete time steps (clock edges), updating the membrane potential at each step based on incoming spikes. However, using operational amplifiers (op-amps) results in significant area and power overheads, especially given the large number of neurons in neural models. Additionally, high sparsity is a characteristic of event-based neural models. 
To address these challenges, we propose utilizing a pair of storage cells (i.e., capacitance) within each neuron engine (\NE), each referred to as a virtual neuron. This approach allows each \NE to model multiple neurons using a single circuit for integration and firing, leading to substantial reductions in energy consumption and area requirements. 
It is important to note that neurons from a given model should be mapped onto an \NE with lower overhead. We employ an ILP formulation to facilitate this mapping process, taking into account the overlap between neuron operations based on profiles generated through simulations conducted with SNNTorch \cite{ref33}.


In this structure, for each set of synaptic connections, the previously determined voltage of a neuron (which is stored in a dedicated capacitor within the \NE) is restored and applied to the output of the op-amp. Following the integration phase, the resulting output voltage is temporarily stored in the capacitor when the input is accumulated into a voltage. This approach ensures that the stored voltage is preserved for subsequent processing cycles.
To emulate the leaky behavior of the neuron, a portion of the stored voltage in the capacitors is discharged at each time step. The controller generates the command for this discharge process.


\subsection{\Syn Engine}

C2C ladders can act as an analog multiplier, where one of its inputs is analog voltage ($V_{ref}$) while the other one is a multi-bit digital input ($W$) \cite{ref28}. The output of this multiplier is an analog signal ($V_{out}$) which its value is obtained by 

\begin{equation}
V_{out} = V_{ref} \times \left( \sum_{i=0}^{n-1} \left( W_{i} \times 2^{i-n}\right) \right)
\label{eq:c2c}
\end{equation}

\begin{figure}[ht]
    \centering
    \includegraphics[width=\linewidth]{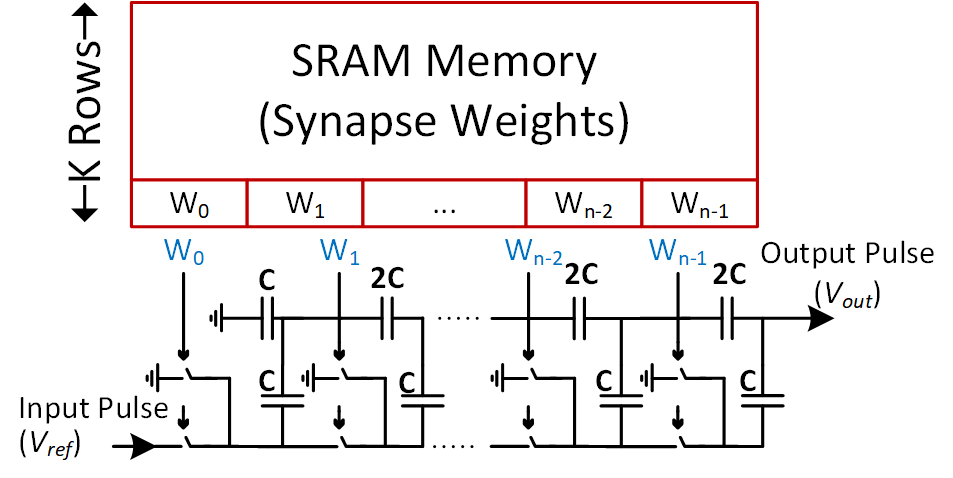} 
    \vspace{-2 em}
    \caption{The proposed \Syn structure}
    \label{fig:Synapse} 
    \vspace{-1 em}
\end{figure}

The digital input of the C2C can be sourced directly from the memory array, allowing it to function as a computing engine located close to the memory array. This arrangement is advantageous for implementing charge-based, CMOS-compatible in-memory computing. 
In the proposed \Syn engine, depicted in \myfigref{fig:Synapse}, a C2C ladder is utilized to model the synaptic connections, with the weights of these connections stored in SRAM memory. The bitlines of the SRAM are connected to the digital input of the C2C ladder. Additionally, Metal-Oxide-Metal (MOM) capacitors can implement the C2C ladder, significantly reducing the area overhead typically associated with C2C capacitance \cite{ref34}.


\subsection{Memory Construction}

\begin{figure*}[ht]
    \centering
    \includegraphics[width=0.8\linewidth]{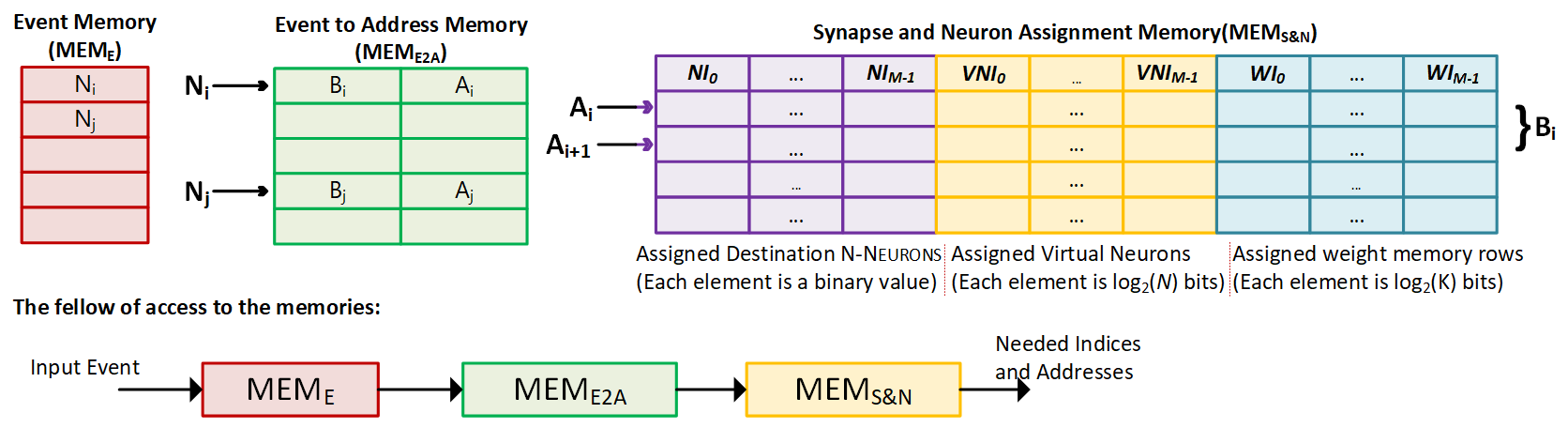} 
    \caption{The memory generated by the ILP problem to input to the system verilog}
    \label{fig:memory} 
\end{figure*}

The memory structures for distributing the pulses of the received events to the proper \Syn and \NE engines in the \PE are shown in \myfigref{fig:memory}.
In \memone, $N_i$ represents the index of a source neuron in the previous layer. This index is also the address for memory access to \memtwo. Note that the number of rows of \memtwo is higher than the number of neurons in the previous layer. Each row of \memtwo has two columns, i.e., $B_i$ and $A_i$. $B_i$ indicates the number of rows in \memthree that are linked to $N_i$, starting from the address pointed by $A_i$. 
\memthree contains information about the synaptic connections and the destination neurons of the $N_i$. 
This memory has three column groups (which are appropriately color-coded). The first column group contains $M$ columns where $M$ denotes the number of the \NE engines. Each of these columns is a binary value ($NI_i$) that indicates to which \NE the received spike should be sent. The second column group contains the virtual neuron indices and identifies the virtual neuron in each \NE to which the destination neuron is mapped. Thus the width of each column of this column type is $log(N)$. Each column of the third and final column group shows the address of the synaptic weight in the weight memory of the \Syn.
Since a source neuron may be connected to more than $M$ available {\NE}s, its connections may be defined in a couple of rows, which, as mentioned before, is indicated by the $B_\star$.


\subsection{Proposed ILP-based Mapping}

The objective of the ILP formulation is to allocate each neuron in the destination layer to a designated capacitor in a \NE for each time step and layer. Once all the connections for a given neuron in the destination layer are processed, the capacitor tied to that neuron must be reassigned to another. This problem is NP-complete and can be modeled as an ILP problem. In the following paragraphs, we provide details of the ILP formulation for assigning synaptic connections and neurons to \PE and \NE in a spiking neural network, ensuring compliance with several imposed constraints. Note that this ILP must be solved for each layer individually, requiring multiple ILPs to be solved at each time step. Below, we present the ILP formulation for a single layer.

ILP is particularly valuable in this context because it allows us to define the objective function and constraints clearly and comprehensively. By employing the ILP formulation, we can minimize the number of unassigned neurons in the spiking network while ensuring efficient hardware utilization. The constraints guarantee that the capacity of each \NE, as defined by the number of capacitors assigned to it, is not exceeded. Additionally, assigning neurons in the spiking network to capacitors within the NE is done in a manner that adheres to fan-out limitations and minimizes the communication overhead.

The binary decision variable \( x_{i,j,k} \), as defined in \myeqref{eq:xij}, is set to 1 if neuron \( i \) in the destination layer is assigned to the \( k^{th} \) capacitor of \( j^{th} \) \NE and 0 otherwise. This variable is crucial to the optimization problem, as it reflects the assignment of neurons to capacitors in {\NE}s.

\begin{equation}
    x_{i,j,k} = 
    \begin{cases} 
    1, & \parbox[t]{.6\linewidth}{\centering if neuron \( i \) is assigned to the \( k^{th} \) capacitor of the \( j^{th} \) \NE} \\
    0, & \parbox[t]{.6\linewidth}{\centering otherwise}
    \end{cases}
\label{eq:xij}
\end{equation}

The objective of the ILP is to minimize the \myeqref{eq:obj}, which is the total number of unassigned neurons. This is formulated by minimizing the sum of \( (1 - x_{i,j,k}) \) over all possible combinations. By doing so, the ILP ensures that as many neurons as possible are assigned to each \NE, reducing the number of unassigned neurons. Here $N_1$ shows the number of neurons in the destination layer, $M$ denotes the number of \NE inside \PE, and $N$ shows the number of capacitors inside each \NE.

\begin{equation}
\text{Minimize} \quad \sum_{i=1}^{N_1} \sum_{j=1}^{M} \sum_{k=1}^{N} (1 - x_{i,j,k})
    \label{eq:obj}
\end{equation}

The first constraint stated in \myeqref{eq:firstcon}, called the Engine Capacity Constraint, ensures that the total number of neurons assigned to any \NE does not exceed the number of capacitors in that \NE, denoted by \( N \). For each \NE, the sum of \( x_{i,j,k} \) over all neurons \( i \) and capacitors \(k\) must be less than or equal to \( N \). This constraint guarantees that the capacity limitations of each \NE are respected, preventing overloading. This is shown in \myeqref{eq:firstcon}.

\begin{equation}
    \sum_{i=1}^{N_1} \sum_{k=1}^{N} x_{i,j,k} \leq N, \quad \forall j \in \{1, 2, \ldots, M\}
    \label{eq:firstcon}
\end{equation}

The second constraint given in \myeqref{eq:secondcon}, i.e., the Unique Engine Assignment Constraint, requires that each neuron in the destination layer is assigned to exactly one \NE. This is achieved by ensuring that the sum of \( x_{i,j,k} \) over all {\NE}s and capacitors equals 1 for each neuron. This constraint ensures that each neuron has a unique assignment, preventing conflicts and ensuring that data routing is straightforward and efficient.

\begin{equation}
    \sum_{j=1}^{M} \sum_{k=1}^{N} x_{i,j,k} = 1, \quad \forall i \in \{1, 2, \ldots, N_1\}
    \label{eq:secondcon}
\end{equation}

The third constraint provided in \myeqref{eq:thirdcon}, i.e., the Connection Constraint in \myeqref{eq:thirdcon}, applies to neurons in the source layer. It states that the sum of connections from each neuron in the source layer to neurons in the destination layer, across all {\NE}s in a \PE, must not exceed the fan-out limit for that neuron. For each neuron \( m \) in the source layer, the sum of \( x_{i,j,k} \) for all connected neurons \( i \) and all {\NE}s and capacitors must be less than or equal to the predefined fan-out limit for that neuron. This constraint helps maintain the structural integrity of the network, ensuring that the connections do not exceed the maximum allowable fan-out. In this equation, $N_2$ denotes the number of neurons in the source layer, and $S_m$ is the set of connections starting from the source layer to $i^{th}$ neuron in the destination layer.

\begin{equation}
\sum_{i \in S_m} \sum_{j=1}^{M} \sum_{k=1}^{N} x_{i,j,k} \leq \text{fanout}_m, \quad \forall m \in \{1, 2, \ldots, N_2\}
     \label{eq:thirdcon}
\end{equation}

Together, these constraints and the objective function create a robust framework for optimizing the assignment of neurons to {\NE}s and capacitors in them in a way that is both efficient and adheres to hardware limitations. 

\section{Results and Discussion}

\subsection{Experimental Setup}

In this study, we employed the SNNTorch framework, a Python-based library for building and training spiking neural networks, to execute our models on two neuromorphic datasets, N-MNIST \cite{ref35} and CIFAR10-DVS \cite{ref36}. These datasets represent spike-based versions of the MNIST and CIFAR10 datasets, respectively, and are frequently used in neuromorphic computing research due to their suitability for testing SNN models.

In this work, we have considered a multi-layer perception (MLP) neural model for both datasets. The network architectures were 200/100/40/10 and 1000/500/200/100/10, in the case of the N-MNIST and CIFAR10-DVS, respectively. Also, we designed two proposed accelerator models with 4 (Accel\textsubscript{1}) and 5 (Accel\textsubscript{2}) {\PE}s for executing the N-MNIST and CIFAR10-DVS. 
The total weight memory in each \PE of the Accel\textsubscript{1} (Accel\textsubscript{2}) was 400 KB (20 MB), and we have considered 10 (20) \NE inside each \PE. Also, each \NE consisted of 16 (32) virtual neuron.

For both datasets, pruning and post-training quantization techniques were applied to optimize the networks while preserving accuracy. For the N-MNIST (CIFAR10-DVS) dataset, the network achieved an accuracy of 94.75\% (65.38\%) before pruning. After applying the unstructured L1 pruning and 8-bit quantization, the accuracy slightly dropped to 94.1\% (65.03\%). \mytabref{tab:comparison} summarizes the details of the models and their training parameters.

\begin{table}[ht]
\vspace{-2 em}
\centering
\renewcommand{\arraystretch}{1.2} 
\caption{Details of the models and their training parameters}
\scriptsize
\begin{tabular}{|>{\centering\arraybackslash}m{3cm}|>{\centering\arraybackslash}m{2.2cm}|>{\centering\arraybackslash}m{2.2cm}|}
\hline
\textbf{Attribute} & \textbf{N-MNIST} & \textbf{CIFAR10-DVS} \\
\hline
\textbf{Number of Parameters} & 0.49 M & 33.4 M\\
\hline
\textbf{Number of Hidden Layers} & 3 (200/100/40) & 4 (1000/500/200/100) \\
\hline
\textbf{Output Neurons} & 10 & 10 \\
\hline
\textbf{Learning Rate} & 1e-3 & 1e-3 \\
\hline
\textbf{Epochs} & 50 & 100 \\
\hline
\textbf{Pruning Technique} & L1 pruning & L1 pruning \\
\hline
\textbf{Quantization} & 8-bit post-training quantization & 8-bit post-training quantization \\
\hline
\end{tabular}
\label{tab:comparison}
\end{table}

\begin{table*}[ht]
\centering
\renewcommand{\arraystretch}{1.2}
\caption{Comparison of Our Proposed Accelerator with Prior Work Based on Various Features}
\begin{tabularx}{\linewidth}{|>{\raggedright\arraybackslash}X|c|c|c|c|c|c|c|}
\hline
\textbf{Author} & \textbf{Neural Operations} & \textbf{TOPS/Watt} & \textbf{Bit Width} & \textbf{Technology} & \textbf{Evaluated Dataset} & \textbf{\# Neurons} \\ 
\hline
\textbf{\Menage (Accel\textsubscript{1})} &  Analog LIF & 3.4 & 8 & 90nm & N-MNIST & 40\\ 
\hline
\textbf{\Menage (Accel\textsubscript{2})} & Analog LIF & 12.1 & 8 & 90nm & CIFAR10-DVS & 100 \\ 
\hline
\textbf{Liu et al. 2023 \cite{ref26}} &  Mixed Signal LIF & 1.88 & 4 & 180nm & MIT-BIH Arrhythmia & 102 \\ 
\hline
\textbf{Qi et al. 2024 \cite{ref37}} &  Mixed Signal LIF & 0.67–5.4 & 8 & 55nm & N/A & 128-256\\ 
\hline
\textbf{Zhang et al. 2024 \cite{ref38}}  & Digital LIF & 0.66 & 8-10 & 28nm & N-MNIST, DVS-Gesture, N-TIDIGIT, SeNic & 522 \\ 
\hline
\textbf{Liu et al. 2024 \cite{ref39}} & Digital LIF & 0.26 & N/A & 22nm & N-MNIST, DVS-Gesture & N/A \\ 
\hline
\end{tabularx}
\label{tab:comp1}
\end{table*}

For the mapping process, the ILP was implemented in Python, and the PuLP package was utilized to solve the ILP problem. 
Also, the digital part of the accelerator was developed by SystemVerilog HDL, and its design parameters were extracted using the Synopsys Design Compiler tool. On the other hand, its analog parts have been defined by Spice scripts and simulated by Synopsys HSpice tool. 

\subsection{Results}
\begin{figure}[ht]
    \centering
    \includegraphics[width=0.8\linewidth]{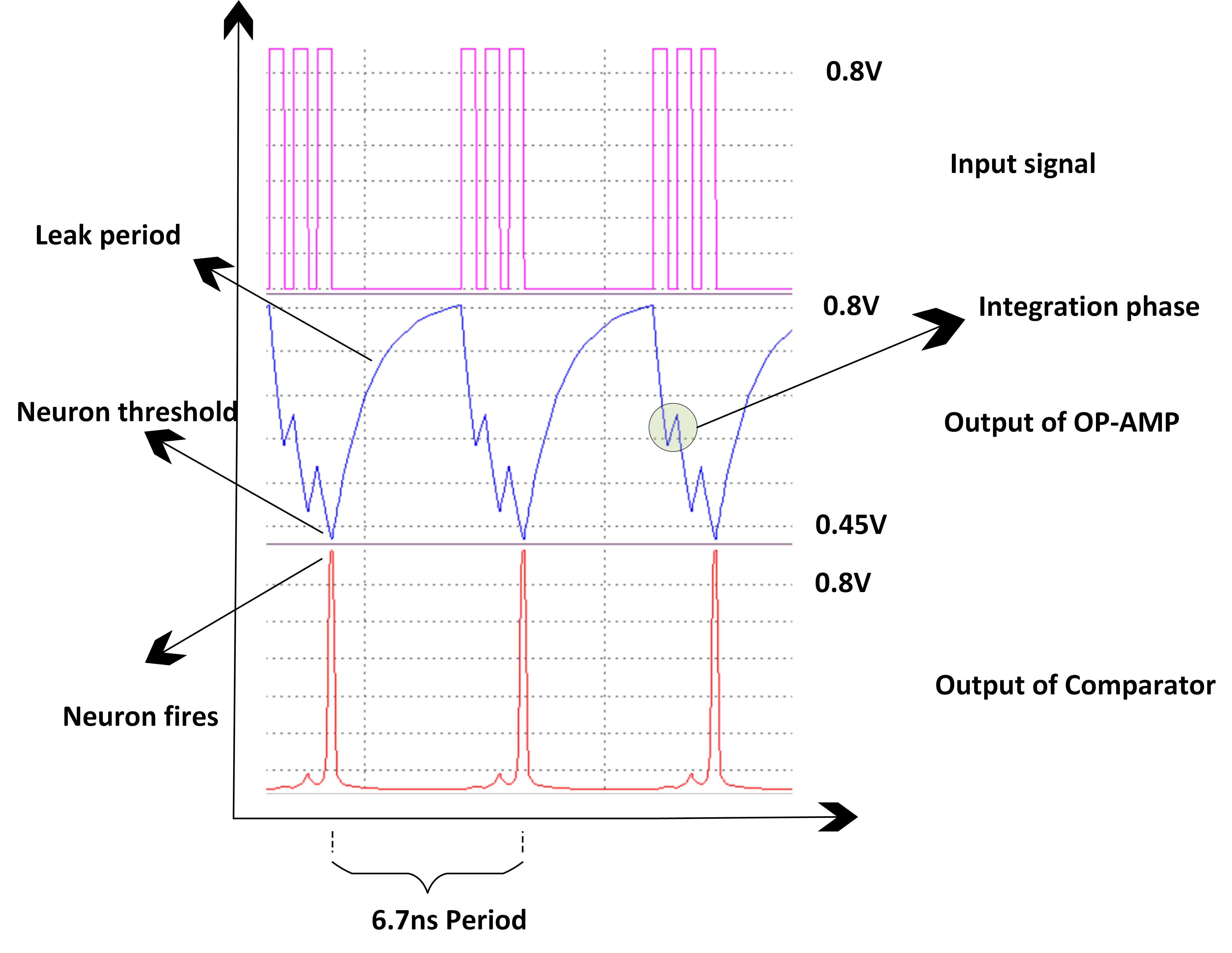} 
    \caption{Spice simulation of the designed \NE circuit with input, output, and integration voltage}
    \label{fig:Neuron} 
\end{figure}

The functionality of the \NE is shown in \myfigref{fig:Neuron}.
The output spike of the \NE, which is the output of the comparator (second op-amp), and the output of the first op-amp in the neuron circuit, along with the input signal, is shown in this figure. The power consumption and delay for each \NE are 97nW and 6.72 ns, respectively. 
Also, the whole \PE simulation shows that the operating frequency of the system is 103.2MHz, and Accel\textsubscript{1} (Accel\textsubscript{2}) could provide 3.4 (12.1) TOPS/W energy efficiency. 


\begin{figure}[ht]
    \centering
    \includegraphics[width=0.8\linewidth]{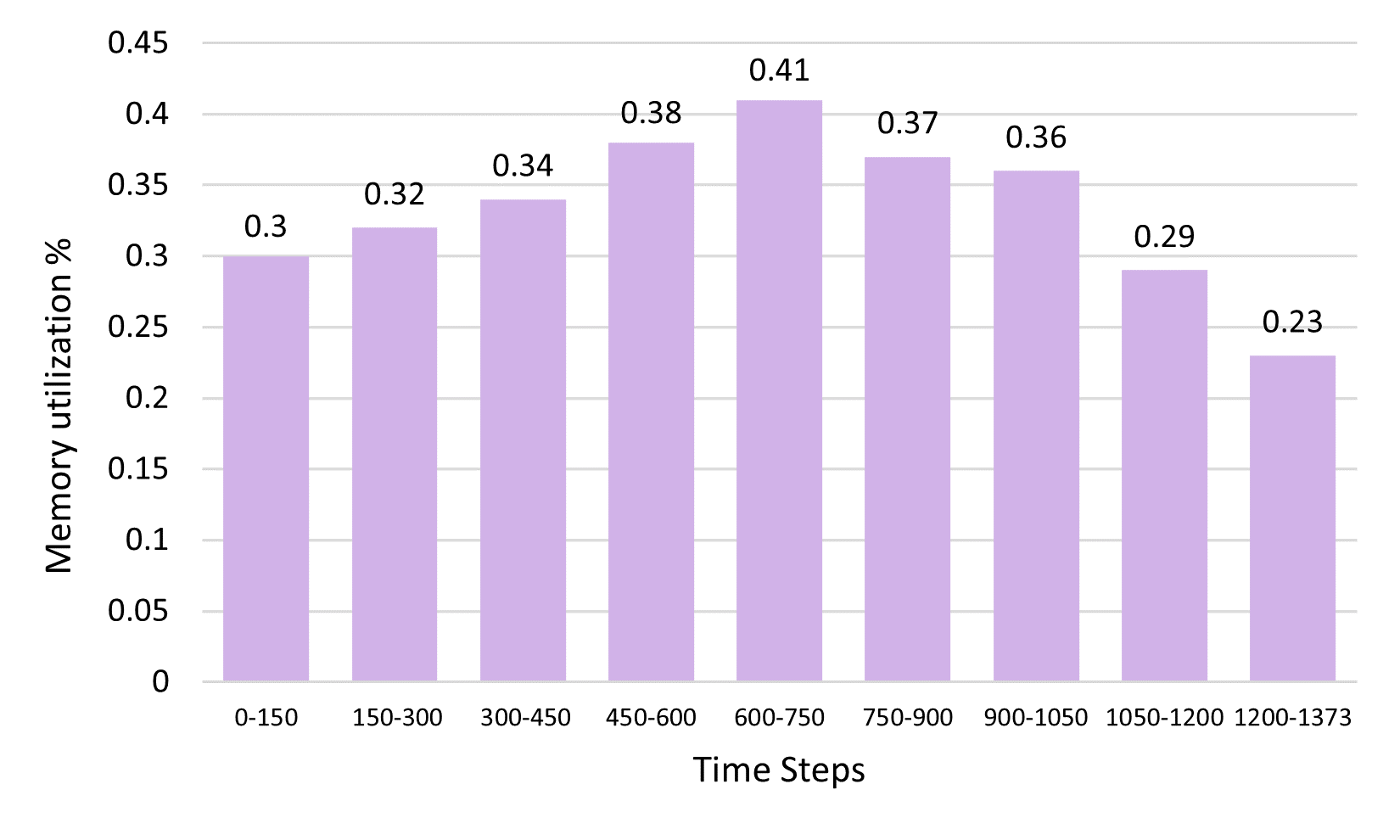} 
    \vspace{-1 em}
    \caption{The memory utilization of running N-MNIST on Accel\textsubscript{1} in different time steps.}
    \label{fig:mem1} 
\end{figure}

\begin{figure}[ht]
    \centering
    \includegraphics[width=0.8\linewidth]{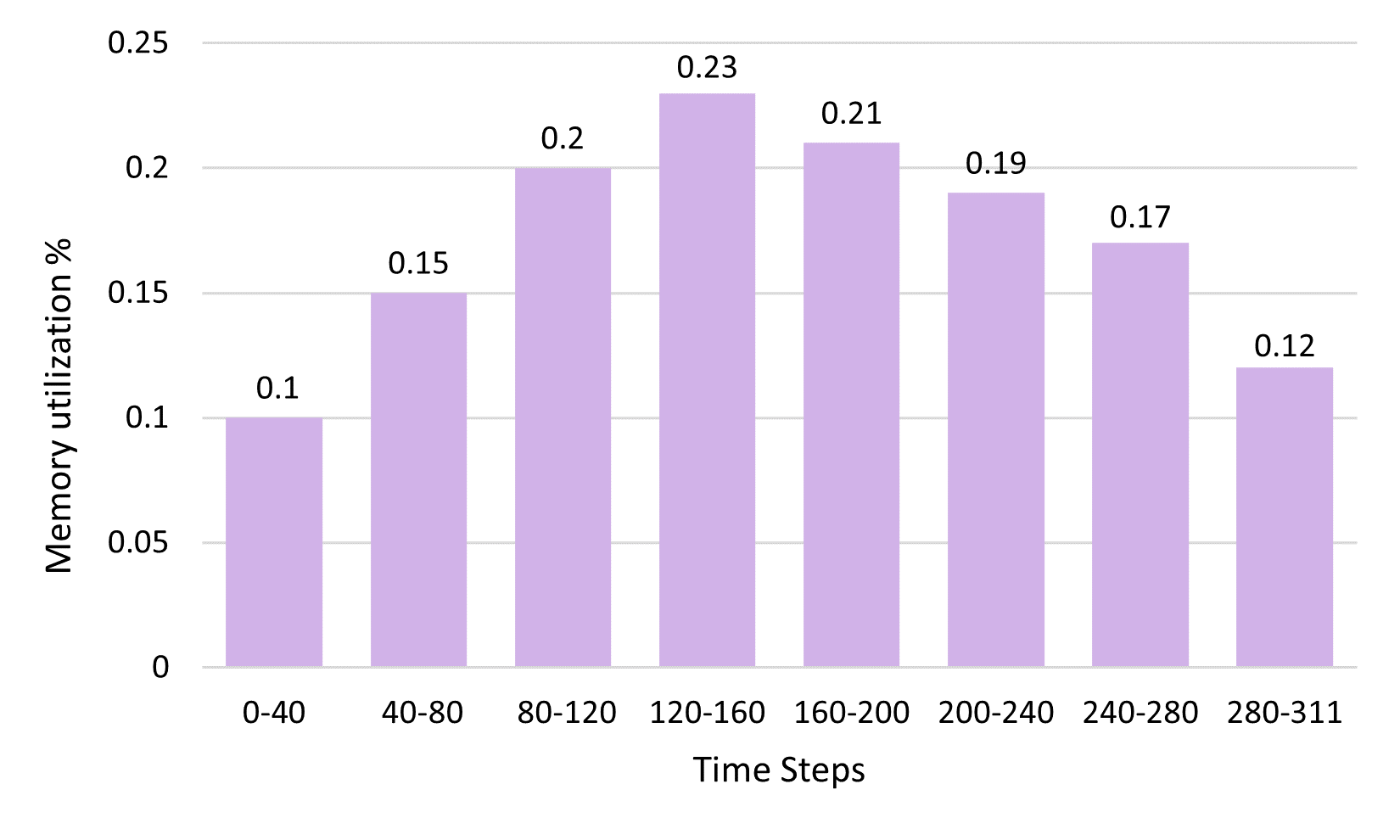} 
    \vspace{-1 em}
    \caption{The memory utilization of running CIFAR10-DVS on Accel\textsubscript{2} in different time steps.}
    
    \label{fig:mem2} 
\end{figure}

\myfigref{fig:mem1} and \myfigref{fig:mem2} illustrates the average memory usage of \memthree, the highest memory-consuming component in this study, on N-MNIST and CIFAR10-DVS datasets using the proposed Accel\textsubscript{1} and Accel\textsubscript{2}, respectively, at various time steps during the processing of a single input image. The figures highlight that, due to the high sparsity of spikes in these datasets, average memory usage remains relatively low most of the time. However, there are instances where memory usage rises at specific time steps or in certain layers, particularly when a large number of spikes occur simultaneously, which the architecture efficiently manages. Notably, CIFAR10-DVS exhibits higher spike activity, leading to increased memory usage compared to N-MNIST.
Finally, \mytabref{tab:comp1} compares the features of our proposed neuromorphic accelerator with those of some fully programmable prior work. This table shows that our architecture is more power efficient compared to the previous work. Additionally, our approach employs substantially fewer neurons than other methods, even when handling a more complex dataset. This reduction in neuron count contributes to a significant decrease in the overall area, enhancing the efficiency of the design.

\section{Conclusion}
This work introduced a mixed-signal neuromorphic accelerator architecture that enhances the efficiency of event-based neural models using analog computing with C2C ladders and LIF for synapse and neuron emulation, respectively. By leveraging virtual neurons, the design improves resource utilization and power efficiency. 
The spike management inside the proposed \PE has been done through soft codes, which have been extracted during the model compilation. 
For efficient mapping of the neural model on the proposed accelerator, we have suggested an ILP formulation.
The energy efficiency of the designed accelerator for executing the CIFAR10-DVS was 12.1 TOPS/W.

\bibliographystyle{unsrt}
\bibliography{ref}


\end{document}